\documentclass[aps,prb,groupedaddress,twocolumn]{revtex4-1}

\usepackage[fleqn]{amsmath}
\usepackage{hyperref}
\usepackage{graphicx}
\usepackage{amssymb}
\usepackage[usenames,dvipsnames]{xcolor}
\usepackage[english]{babel}

\DeclareMathOperator{\Tr}{Tr}

\newcommand{\rs}[2]{\ensuremath{\textcolor{red}{\sigma_{#1}^{#2}}}}
\newcommand{\bt}[2]{\ensuremath{\textcolor{blue}{\tau_{#1}^{#2}}}}
\newcommand{\zz}{\ensuremath{{Z}_2}}
\newcommand{\dd}{\ensuremath{{Z}_2\times{Z}_2}}

\newcommand{\bigO}[1]{\ensuremath{\mathcal{O}(#1)}}
\hypersetup{colorlinks, linkcolor={blue},citecolor={red}}

\def\inlinecite{Ref.~\onlinecite}
\def\inlinecites{Refs.~\onlinecite}

\newcommand{\bra}[1]{\mathinner{\langle{#1}|}}
\newcommand{\ket}[1]{\mathinner{|{#1}\rangle}}

\newcommand{\ketbra}[1]{\ensuremath{\ket{#1}\hspace{-.25em}\bra{#1}}}

\allowdisplaybreaks

\begin{document}
  \title{Multiscale Entanglement Renormalisation Ansatz for Spin Chains with Continuously Varying Criticality}
  \author{Jacob C. Bridgeman}
  \affiliation{Centre for Engineered Quantum Systems, School of Physics, The University of Sydney, Sydney NSW 2006, Australia}
  \author{Aroon O\rq{}Brien}
  \affiliation{Centre for Engineered Quantum Systems, School of Physics, The University of Sydney, Sydney NSW 2006, Australia}
  \author{Stephen D. Bartlett}
  \affiliation{Centre for Engineered Quantum Systems, School of Physics, The University of Sydney, Sydney NSW 2006, Australia}
  \author{Andrew C. Doherty}
  \affiliation{Centre for Engineered Quantum Systems, School of Physics, The University of Sydney, Sydney NSW 2006, Australia}

  \begin{abstract}
    We use the multiscale entanglement renormalisation ansatz (MERA) to numerically investigate three critical quantum spin chains with $\dd $ on-site symmetry: a staggered XXZ model, a transverse field cluster model, and the quantum Ashkin-Teller model.  All three models possess a continuous one-parameter family of critical points.  Along this critical line, the thermodynamic limit of these models is expected to be described by classes of $c=1$ conformal field theories (CFTs) of two possible types:  the $S^1$ free boson and its $\zz $-orbifold.  Our numerics using MERA with explicitly enforced $\dd $ symmetry allow us to extract conformal data for each model, with strong evidence supporting the identification of the staggered XXZ model and critical transverse field cluster model with the $S^1$ boson CFT, and the Ashkin-Teller model with the $\zz $-orbifold boson CFT.  Our first two models describe the phase transitions between symmetry protected topologically ordered phases and trivial phases, which lie outside the usual Landau-Ginsburg-Wilson paradigm of symmetry breaking.   Our results show that a range of critical theories can arise at the boundary of a single symmetry protected phase.
  \end{abstract}

  \maketitle
  \section{Introduction}\label{sec:Introduction}
  Quantum spin models are an area of extensive theoretical and numerical study, due to their relative simplicity and wide descriptive power.  Simple spin models can exhibit a variety of exotic ground states properties, such as topological order\cite{WenTopoReview} and symmetry-protected topological order\cite{ChenGuLiuWen}.  Studying the critical behaviour that describes a system at the transition point between two quantum phases can also be investigated with quantum spin models but brings new challenges. At these critical points, the models become gapless and many of the exact results that have been proven for gapped systems break down.  In particular, for critical systems the correlation length in the ground state diverges and the area law for entanglement entropy is violated\cite{Eisert2010}.  In the thermodynamic limit, the behaviour at the critical point is described by a conformal field theory (CFT)\cite{Ginsparg1988,DiFrancesco1997}.

  A wide range of numerical methods have been developed to study the behaviour of quantum spin chains, including their critical behaviour, despite the difficulties in working with a Hilbert space that grows exponentially in the chain length.  Numerical methods based on tensor networks\cite{Verstraete2008,Orus2014} have been highly successful in recent years as an efficient method for studying a wide range of spin lattice models. In particular, for one dimensional gapped systems, methods based on matrix product states are known to efficiently describe ground state properties\cite{Landau2014}.  However, this behaviour does not generalise to gapless systems \cite{Verstraete2006} (although c.f.~\inlinecite{Stojevic2014}). For a tensor network description to naturally describe a critical spin chain, is should capture the area law violation, and one such description is the Multiscale Entanglement Renormalisation Ansatz \cite{Vidal2007} (MERA). The MERA has previously been shown to accurately and efficiently reproduce conformal data of critical spin chains\cite{Pfeifer2009,Evenbly2010c,Giovannetti2008}.

  In this paper, we use the MERA to numerically study three related spin models along a line of criticality: a staggered XXZ model, a transverse field cluster model\cite{Doherty2009}, and the quantum Ashkin-Teller model\cite{Solyom1981,Ashkin1943}. These models are of interest because they have critical exponents that vary continuously as a function of model parameters.  All three models possess an on-site $\dd $ symmetry that will play an important role in our analysis.

  For the first two models of interest, the critical line corresponds to a phase transition between a phase possessing nontrivial symmetry protected topological order (SPTO) for this symmetry group\cite{Chen2011} and a trivial phase. Such systems are also of particular interest due to their connection to measurement-based quantum computation \cite{Doherty2009,Else2012,Else2012NJP}. There has been extensive investigation into symmetry protected topological phases, but only a few studies have investigated transitions out of them\cite{Chen2013248,Lu2014}. Such phase transitions are not part of the usual Landau-Ginsburg-Wilson paradigm of phase transitions because there is no broken symmetry on either side of the transition. Despite the fact that the models we consider here involve a transition between a fixed SPTO phase and the trivial phase, the boundary is described by a range of critical theories.

  In contrast, the Ashkin-Teller model possesses a conventional symmetry broken phase and for that model the phase transition of interest is between the symmetry broken and trivial phases. Again, a range of critical theories describe the phase boundary.

  The thermodynamic limit of these three models at their critical points are described by a special class of CFT, namely those with central charge $c=1$. These CFTs do not correspond to any member of the series of so-called {\it minimal models} that have been the main focus of numerical research that has used MERA to study CFTs\cite{Pfeifer2009,Giovannetti2008}. This class of CFT is characterised by a renormalisation group with an exactly marginal field, leading to a single parameter family of theories that reproduces the continuously varying critical exponents of the underlying spin models.  By varying the coupling in the spin models, we can tune the system along this line of criticality. This continuous variation of scaling dimensions is a distinctive feature of $c=1$ CFTs\cite{Cardy1987}.

  Here, we show that the MERA allows us to extract conformal data from all three models described above possessing a continuous line of criticality, and show that these data agree with the expected $c=1$ CFTs including the variation of the conformal data along the line of criticality.  Specifically, the staggered XXZ model and the transverse field cluster model are both shown to be consistent with the $c=1$ $S^1$ free boson CFT, whereas the quantum Ashkin-Teller model gives data consistent with the $c=1$ orbifold CFT.  Our results provide further evidence of the usefulness of MERA in describing critical systems in one dimension.

  This paper is structured as follows.  In Section~\ref{sec:CFTs}, we review CFTs as they pertain to the three spin models studied. In Section~\ref{sec:MERA} we review the details of the MERA algorithm. In Section~\ref{sec:Results} we present results from the numeric simulations for each model, and compare with the expected CFT data.

  \section{$c=1$ Boson CFTs and associated spin models}\label{sec:CFTs}

  We will consider three different one dimensional quantum spin models at their respective critical points. In each of these models there is a line of critical points along which the critical exponents vary continuously. These critical points are believed to be described by certain conformal field theories with marginal operators that generate flow along these lines of criticality. In this section, we briefly review these conformal field theories and introduce our three spin chain models.

  \subsection{$c=1$ CFTs}

  \begin{figure}
    \includegraphics{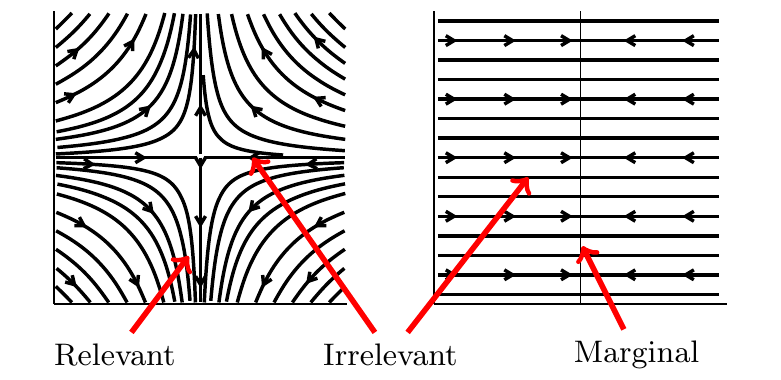}
    \vspace{5mm}
    \caption{Illustrative phase space diagrams and renormalisation group (RG) flows for models with a unique RG fixed point (left) and a line of criticality (right).  RG relevant operators have $\Delta<2$ and grow under renormalization, whilst those with $\Delta>2$ are called irrelevant, since a CFT deformed by such an operator flow back to the fixed point. Deformations by an exactly marginal operator lead to a new conformal fixed point. As the size of the deformation is varied, the conformal dimensions vary continuously. }\label{fig:Marginal}
  \end{figure}

  Conformal field theories (CFTs) are field theories that are invariant under all conformal (angle preserving) space-time transformations. They describe the thermodynamic limit of critical lattice models~\cite{DiFrancesco1997,Ginsparg1988}. Due to the abundant symmetry, a (1+1)D CFT is completely specified by knowledge of (\textit{i}) its central charge $c$; (\textit{ii}) the primary fields $\phi$ and their scaling dimensions $(h,\bar{h})$, which are eigenoperators of the scaling transformation with eigenvalues $\Delta=h+\bar{h}$; and (\textit{iii}) the operator product expansion coefficients for these fields. It is the primary field scaling dimensions $\Delta$ that predict the critical exponents of the associated critical lattice model.

  We focus our attention on CFTs with $c=1$, which mainly fall into two categories:  the free boson theory and its $\zz $-orbifold\cite{DiFrancesco1997,Ginsparg1988}.  These $c=1$ boson CFTs are characterised by the presence of an exactly marginal primary field, with scaling dimension 2, meaning it is fixed under rescaling. Perturbing a CFT by an exactly marginal term gives another CFT, leading to continuously varying families of theories (see Figure 1). This phenomenon is not seen in unitary CFTs with $c<1$. These two continuous families of CFTs will correspond to the critical lines in our various quantum spin chains.

  The first relevant family of CFTs, the \textit{$S^1$ boson} CFT,  is the theory of a massless free boson $\varphi(x)$. We will be interested in periodic boundary conditions and we will choose a parameterisation such that $x\in[0,2\pi)$. We will focus on a compactified version of the free boson where $\phi$ itself takes values on a circle of radius $R_C$, so that
  $\varphi(x)\equiv\varphi(x)+2\pi R_C$.
  As a result when we move around a circle in the $x$ co-ordinate, $\varphi$ does not need to be strictly periodic but may twist around $m$  times so that $\varphi(x+2\pi)=\varphi(x)+2\pi mR_C$.  The primary field scaling dimensions are known for this theory and generally have a non-trivial dependence on $R_C$. The free boson has a natural  internal $SO(2)$ symmetry given by the translation $\varphi(x)\equiv\varphi(x)+\theta R_C$, where $\theta \in [0,2\pi)$, as well as a $\zz $ symmetry given by $\varphi(x)\rightarrow -\varphi(x)$.

  The \textit{orbifold} boson CFT once again involves a bosonic field $\varphi(x)$ compactified on a circle of radius $R_O$, so that
  $\varphi(x)\equiv\varphi(x)+2\pi R_O$~\cite{DiFrancesco1997,Ginsparg1988}. The theory has two sectors. The first corresponds to the subspace of symmetric states of the free boson that map to themselves under $\varphi(x)\rightarrow -\varphi(x)$. The second sector corresponds to quantising the free boson with ``twisted" boundary conditions for which $\varphi(x+2\pi)=-\varphi(x)$ and projecting out the states that are symmetric under $\varphi(x)\rightarrow -\varphi(x)$~\cite{DiFrancesco1997,Ginsparg1988}. A characteristic of the twisted sector of the orbifold boson is that it contains a number of primary fields whose scaling dimension is independent of $R_O$. Due to this construction the $O(2)$ symmetry of the free boson model is broken down to a $\zz $ symmetry given by $\varphi(x)\rightarrow \varphi(x)+\pi R_O$, and the model retains its symmetry under $\varphi(x)\rightarrow -\varphi(x)$ so that the natural symmetry group is $\dd $.

  For a more detailed examination of these CFTs, including expressions for their spectra of primary fields, we refer to \inlinecite{DiFrancesco1997}.
  \subsection{Spin models at criticality}

  In the following, we present three critical spin models whose thermodynamic limit is described by $c=1$ CFTs.  All three models will be defined on a line with periodic boundary conditions, and will possess an on-site $\dd $ symmetry.

  Each of these models have a line of critical points separating an ordered phase from a disordered phase. For two of the models, the staggered XXZ model and the transverse field cluster model, the ordered phase displays symmetry protected topological order\cite{Furukawa2012,Else2012}.  In contrast, for the quantum Ashkin-Teller model the ordered phase displays conventional symmetry breaking of the $\dd $ symmetry.

\begin{figure}
    \begin{center}
      \includegraphics{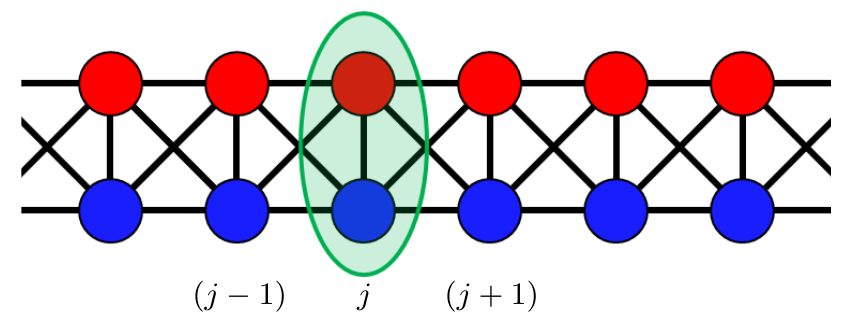}
      \caption{Each of the spin models is defined on a pair of parallel chains. The $\rs{}{}$ operators act only on the red (top) chain and the $\bt{}{}$ act only on the blue (bottom). The green region indicates a single site. Each of the models is then defined by a 1D nearest-neighbour Hamiltonian.
      }\label{fig:ATChains}
    \end{center}
  \end{figure}

  To clarify the on-site nature of this symmetry, we will assign two spin-$1/2$ particles to each site, and colour them red and blue (Fig. \ref{fig:ATChains}).  At site $j$, the Pauli spin operators for each spin are denoted $\rs{j}{\alpha}$ and $\bt{j}{\beta}$ for $\alpha, \beta =X,Y,Z$.

  \subsubsection{Staggered XXZ Model}

  The \textit{staggered XXZ model} is described by the Hamiltonian
  \begin{multline}
    H_{\rm{XXZ}}=-\sum_{j=1}^N\left(\rs{j}{X}\bt{j}{X}+\rs{j}{Y}\bt{j}{Y}-\lambda \rs{j}{Z}\bt{j}{Z}\right)\\
    -\beta\sum_{j=1}^{N-1}\left(\bt{j}{X} \rs{j+1}{X}+\bt{j}{Y}\rs{j+1}{Y}-\lambda \bt{j}{Z}\rs{j+1}{Z}\right).
  \end{multline}
  Note that this model has an $O(2)$ symmetry group, corresponding to rotating each spin about the $z$-axis and a $\zz $ symmetry corresponding to flipping each spin.

  In this work we will be mainly concerned with a $\dd $ subgroup of these symmetries that will remain on-site in all of our models, as we will incorporate this symmetry explicitly into our numerical MERA simulations.  This symmetry is also important for considerations of symmetry protected topological order\cite{ChenGuLiuWen}. We can choose generators for this subgroup as follows:
  \begin{align}
    \textcolor{red}{S_1}&=\prod_{j=1}^{N}\rs{j}{X}\bt{j}{X}&\textcolor{blue}{S_2}=\prod_{j=1}^{N}\rs{j}{Y}\bt{j}{Y}. \label{eqn:XXZSymmetryOps}
  \end{align}
  The generator $\textcolor{red}{S_1}$ corresponds to the spin-flip symmetry of the staggered-XXZ, and therefore to the $\varphi\rightarrow-\varphi$ symmetry of the free boson model. The generator $\textcolor{blue}{S_2}$ corresponds to rotating each spin by $\pi$ about the $Z$-axis and then flipping the spins. Thus it corresponds to the transformation $\varphi\rightarrow -\varphi+\pi$ on the free boson.

  For $\beta=0,\lambda>-1$ the ground state corresponds to pairing the red and blue spins at each site in a maximally entangled state, and thus has the structure of a product state. For $\beta\to\infty,\lambda>-1$ the ground state of this model pairs each blue spin with the red spin on the site to the right in a maximally entangled state. If the spin chain is defined on open boundary conditions there is a four-fold ground state degeneracy associated with unpaired spins at each end of the chain. This state is in the non-trivial symmetry protected topologically ordered state of $\dd $ symmetry~\cite{ChenGuLiuWen}. These two wave functions are the canonical representatives of the trivial and non-trivial (respectively) symmetry protected topologically ordered phases with this symmetry\cite{AKLT}.

  At the phase transition $\beta=1$ separating these two phases, this model is the well-studied XXZ model and is solvable via a Bethe Ansatz\cite{Alcaraz1988}. This information can be used to identify how different values of $\lambda$ correspond to the parameter $R_C$ that specifies the free boson model. The critical line defined by
  \begin{equation}
    \beta=1,\quad \lambda\in[-\sqrt{2}/2,1),\label{eqn:criticalline}
  \end{equation}
  is known to be described by the $S^1$ free boson\cite{Alcaraz1987} compactified on a circle of radius
  \begin{equation}
    R_{C}^2=\frac{2}{\pi}\cos^{-1}(-\lambda).
    \label{eqn:Rc}
  \end{equation}
  Due to the exact solution, this antiferromagnetic model is used extensively in the study of quantum critical points and critical ground states\cite{Kohmoto1981,Alcaraz1987,Alcaraz1988,Alcaraz1988JPHYS,Yamanaka1995}. We also note that the XXZ model with next-nearest neighbour interactions has recently been studied in the context of SPT phases\cite{Ueda2014}. Moreover the study of phase transitions out of the SPTO phase of $SO(3)$ symmetric systems in one dimension in \inlinecite{Chen2013248} corresponds to the Heisenberg model in one dimension and the value $\lambda=1$ in the free boson theory.

  \subsubsection{Transverse field cluster model}

  Cluster states are highly entangled many body states of spin-half particles that arose in the theory of quantum computing. It is possible to simulate any quantum computation by making only local measurements on cluster states on an appropriate lattice~\cite{Raussendorf2001}. The cluster state corresponding to a linear arrangement of spins is the ground state of the following local Hamiltonian $H=-\sum\sigma_{\mu-1}^Z \sigma_{\mu}^X \sigma_{\mu+1}^Z$.
  Recently, cluster state models consisting of this Hamiltonian with various additional terms have been studied, from the perspective of quantum computing\cite{Doherty2009,Else2012,Else2012NJP}, their phase structure as models with symmetry protected topological (SPT) phases\cite{Else2012,Chen2013}, natural models for the gapless edge physics of 2D SPT models\cite{Chen2013,Santos2014}, and nonequilibrium dynamics arising in these models\cite{Montes2012}.

  We define the \textit{transverse field cluster model} (TFCM) by the Hamiltonian
  \begin{multline}
    H_{\rm{TFCM}}=-\sum_{j=1}^N\left(\rs{j}{X}+\bt{j}{X} +\lambda \rs{j}{X}\bt{j}{X}\right) \\
    -\beta\sum_{j=1}^{N-1}\left(\rs{j}{Z}\bt{j}{X} \rs{j+1}{Z}+\bt{j}{Z}\rs{j+1}{X}\bt{j+1}{Z}\right. \\
    \left.+\lambda \rs{j}{Z}\bt{j}{Y}\rs{j+1}{Y}\bt{j+1}{Z}\right).\label{eqn:TFCMHamiltonian}
  \end{multline}
  This model possesses an on-site $\dd $ symmetry generated by
  \begin{align}
    \textcolor{red}{S_1}&=\prod_{j=1}^{N}\rs{j}{X}&\textcolor{blue}{S_2}=\prod_{j=1}^{N}\bt{j}{X},\label{eqn:TFCMSymmetryOps}
  \end{align}
  and possesses two distinct SPT phases for this symmetry.  For $\beta=0$, the ground state is a product state, i.e., in a trivial phase, whereas if $\beta\to\infty$  the ground state is the well-studied cluster state, known to be within a $\dd $ symmetry protected topological phase\cite{Else2012}.

  The transverse field cluster model can be obtained from the staggered-XXZ model through the transformation
  \begin{align} \label{eqn:ConstructTFCM}
    \rs{j}{X}&\to\rs{j}{X}\bt{j}{Z}&\bt{j}{X}&\to\bt{j}{Z}\\
    \rs{j}{Z}&\to-\rs{j}{Y}\bt{j}{Z}&\bt{j}{Z}&\to\rs{j}{Z}\bt{j}{Y}\nonumber .
  \end{align}
  As this mapping is unitary, the ground state energy and phase transitions of the TFCM model are identical to those of the XXZ model. The mapping (\ref{eqn:ConstructTFCM}) preserves locality, meaning that local operators confined to $k$ sites of the staggered XXZ model map to local operators of the transverse field cluster model confined to the same $k$ sites, and that this transformation can be performed using local unitaries applied to each site.  Finally, this mapping respects the $\dd $ symmetry of the models. As a result the ground states of the two models must be in the same phase\cite{Chen2010}.
  This is one way of seeing that the ordered phase of the transverse field cluster model possesses symmetry protected topological order. Moreover we expect that
  the thermodynamic limit of TFCM on the line defined by Eq.~(\ref{eqn:criticalline}) likewise corresponds to the $S^1$ boson with radius $R_{C}$ determined by $\lambda$ as in Eq.~(\ref{eqn:Rc}).
  Finally the mapping (\ref{eqn:ConstructTFCM}) maps the symmetry operations $\textcolor{red}{S_1},\textcolor{blue}{S_2}$ of the staggered XXZ model to the corresponding symmetry operators for the transverse field cluster model. Thus we can identify these two symmetry operations in terms of the mappings $\varphi\rightarrow -\varphi$ and $\varphi\rightarrow -\varphi +\pi$, respectively, in the free boson model.

  \subsubsection{Quantum Ashkin-Teller model}

  The \textit{quantum Ashkin-Teller} (AT) model\cite{Solyom1981,Ashkin1943} is defined by the Hamiltonian
  \begin{flalign}
    \begin{split}
      H_{\rm{AT}}=&-\sum_{j=1}^N\left(\rs{j}{Z}+\bt{j}{Z}+\lambda \rs{j}{Z}\bt{j}{Z}\right)
      \\&-\beta\sum_{j=1}^{N-1}\left(\rs{j}{X}\rs{j+1}{X}+\bt{j}{X}\bt{j+1}{X}+\lambda \rs{j}{X}\bt{j}{X}\rs{j+1}{X}\bt{j+1}{X}\right).
    \end{split}&&\raisetag{3\baselineskip}
  \end{flalign}
  This model consists of a pair of transverse-field Ising chains $\rs{}{},\bt{}{}$ coupled by two- and four- spin terms. In particular, when $\lambda=0$, we have a decoupled pair of Ising chains.

  This model also possesses an on-site $\dd $ symmetry, generated by
  \begin{align}
    \textcolor{red}{S_1}&=\prod_{j=1}^{N}\rs{j}{Z}&\textcolor{blue}{S_2}=\prod_{j=1}^{N}\bt{j}{Z}.\label{eqn:ATSymmetryOperators}
  \end{align}

  We note that with open boundary conditions this model is also unitarily related to the transverse field cluster model, but now through a \emph{nonlocal} unitary transformation
  \begin{align}
    \rs{j}{X}&\to\rs{j}{Z}&\bt{j}{X}&\to\bt{j}{Z}\label{eqn:ConstructAT}\\
    \rs{j}{Z}&\to\left(\prod_{k=1}^{j-1}\bt{k}{Z}\right)\rs{j}{X}&\bt{j}{Z}&\to\bt{j}{X}\left(\prod_{k=j+1}^{N}\rs{k}{Z}\right)\nonumber.
  \end{align}
  Under this mapping the four-fold degenerate ground space of the transverse field cluster model maps to the four-fold degenerate groundspace of the Ashkin-Teller model. Although this mapping is nonlocal, in the bulk it maps local $\dd $-symmetric terms to local $\dd $-symmetric terms, and thus can be viewed as a generalisation\cite{Else2012b} of the map by Kennedy and Tasaki\cite{Kennedy1992,Kennedy1992a}, mapping the SPTO phase of the transverse field cluster model to the symmetry broken phase of the Ashkin-Teller model.  Note, however, that boundary terms are transformed nontrivially by this map; local boundary terms can map to nonlocal ones.  In particular, periodic boundaries in the XXZ and TFCM models pick up extensive stringlike terms in the AT model.

  (Note that the Ashkin-Teller model also possesses a $\zz $ symmetry associated to swapping $\rs{}{}\leftrightarrow\bt{}{}$, however this does not map to a local symmetry in the other models.)

  Because the models are unitarily related, the AT model will possess the same phase structure and ground state energy as the transverse field cluster and staggered XXZ models (and, in particular, will possess the same line of criticality described by Eq.~(\ref{eqn:criticalline})). However, the phases that arise have quite different properties and the CFT corresponding to the critical line where $\beta=1$ will have a different spectrum of primary fields.  Based on finite size simulations\cite{Baake1987,Baake1987a} and CFT arguments\cite{Yang1987,Yang1987a}, the critical line of the AT model has been identified as the orbifold boson with radius
  \begin{align}
    R_O^2=R_C^{-2},\label{eqn:Ro}
  \end{align}
  where $R_C$ is defined in Eq.~(\ref{eqn:Rc}).

  \section{Scale-invariant MERA with symmetries}\label{sec:MERA}

  Following the work of \inlinecites{Evenbly2009,Evenbly2011}, we have independently developed code to optimise multi-scale entanglement renormalisation ansatz (MERA)\cite{Vidal2007}. This class of tensor networks can be used to efficiently describe ground states of critical quantum lattice models.
  Specifically, we give a brief introduction to the scale invariant ternary MERA, closely following \inlinecite{Evenbly2011}, and describe the adiabatic crawling method which we used to streamline the calculation of critical exponents at various locations along the critical line of our $c=1$ spin chain models.

  \subsection{Elements of the MERA}

  \begin{figure}
    \includegraphics{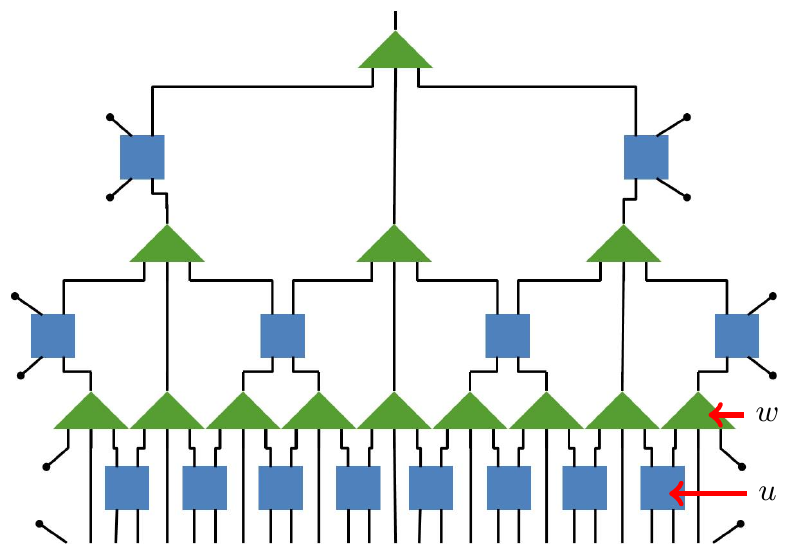}
    \caption{Tensor network diagram for the scale invariant ternary MERA. Isometries are indicated by triangles (green), while disentanglers are rectangles (blue). The layered structure accurately represents a wide range of length scales and captures the scale invariant physics of critical spin chains.}\label{fig:MERA}
  \end{figure}

  The ternary MERA shown in Fig.~\ref{fig:MERA} is constructed from two kinds of tensors: \emph{isometries} $w$
  \begin{align}
    w_\ell:\mathcal{H}_{\ell+1}\to\mathcal{H}_{\ell}^{\otimes3},&&w_\ell^\dagger w_\ell=\mathbb{I}_{\ell+1},\label{eqn:wcond}
  \end{align}
  and unitary \emph{disentanglers} $u$
  \begin{align}
    u_\ell:\mathcal{H}_{\ell}^{\otimes2}\to\mathcal{H}_{\ell}^{\otimes2},&&u_\ell^\dagger u_\ell=\mathbb{I}^{\otimes2}_{\ell},\label{eqn:ucond}
  \end{align}
  where $\mathcal{H}_\ell$ is the Hilbert space of dimension $\chi_\ell$ of one spin (one index) on layer $\ell$.

  Together these tensors perform a real space renormalisation group (RG) transformation, with each layer of the structure corresponding to a description of the model on a different length scale. The free indices correspond to the degrees of freedom associated with the microscopic model of interest, and the isometries map 3 spins on layer $\ell$ onto an effective spin on layer $\ell+1$. The approximation involved in this transformation is controlled by the bond dimension $\chi$; the dimension of the effective spin. If $\chi_{\ell+1}<\chi_\ell^3$, the full microscopic physics cannot be captured and a variational algorithm is used to ensure the low energy subspace is retained. The maximum bond dimension used in the MERA is labelled $\chi$. The unitary tensors rearrange the local Hilbert space, locally removing entanglement and allowing $\chi$ to be relatively small\cite{Vidal2007}.

  Generically, all tensors in the MERA may be different, however for translationally invariant states a single pair of tensors $\{u_{\ell},w_\ell\}$ characterise layer $\ell$. For scale invariant models, such as critical systems, the network becomes simpler still. The description of the model should be scale-invariant, and therefore all layers become identical\cite{Pfeifer2009} and the entire MERA is described by a single pair of tensors $\{u,w\}$, and all bond dimensions are $\chi$.

  The isometries and disentanglers are chosen to minimise the energy of the spin model Hamiltonian using techniques that are described in \inlinecites{Evenbly2009,Evenbly2011}. This results in an iterative algorithm where each step updates either the isometries or the disentanglers and each step can be performed at a cost $\bigO{\chi^8}$, although this can be reduced by modifying the network to include further approximations, such as spatial symmetries\cite{Evenbly2011} or on-site symmetries\cite{Singh2010}.

  \subsection{$\mathcal{G}$-symmetric tensors}

  The incorporation of symmetries can decrease the resources required to optimise the MERA.  If a model possesses an on-site symmetry, such as the $\dd $ symmetry occurring in the models described in section \ref{sec:CFTs}, these internal symmetries may be exploited to further reduce the resource requirement by using \emph{$\mathcal{G}$-symmetric tensors}\cite{Singh2010}.

  A tensor $T_{\alpha_1,\alpha_2,\ldots,\alpha_n}^{\beta_1,\beta_2,\ldots,\beta_m}$ is said to be $\mathcal{G}$-symmetric if it is invariant under the action of the group $\mathcal{G}$ on each index. For example, the requirement for the isometry described above to be $\mathcal{G}$-symmetric is as follows

  \begin{align}
    \begin{array}{l}
      \includegraphics{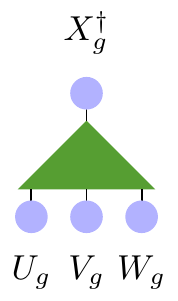}
    \end{array}&=\begin{array}{l}
    \includegraphics{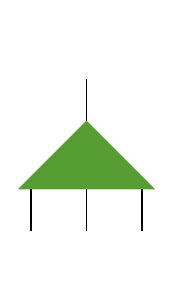}
  \end{array}&\begin{array}{l}\forall\ g\in\mathcal{G},\end{array}\label{eqn:SymmetryCondition}\end{align}
  where $U,V,W,X$ are unitary representations of $\mathcal{G}$.

  By Schur\rq{}s lemma, the tensor decomposes into blocks, where each block transforms as one of the irreducible representations (irreps) of $\mathcal{G}$. In general, due to the block structure of the tensors, we can decompose the indices $\alpha\to(c,d)$, where $c$ labels the irrep and we will call it the charge index, and $d$ is the degeneracy index\cite{Singh2010}. The charge structure is completely specified by the group $\mathcal{G}$. The degrees of freedom for the particular model of interest are completely described by the index $d$.

  In our case we have the group $\mathcal{G}\cong\dd\cong \{ x,y|x^2=y^2=e,xy=yx\}$ whose irreps are given by the character table
  \begin{center}
    \begin{tabular}{|c||c|c|c|c|}
      \hline
      &$e$&$x$&$y$&$xy$\\
      \hline
      \hline
      $D_{(1,1)}$&$1$&$1$&$1$&$1$\\
      \hline
      $D_{(1,{-}1)}$&$1$&$1$&${-}1$&${-}1$\\
      \hline
      $D_{({-}1,1)}$&$1$&${-}1$&$1$&${-}1$\\
      \hline
      $D_{({-}1,{-}1)}$&$1$&${-}1$&${-}1$&$1$\\
      \hline
      \end{tabular}\;,
    \end{center}
    where the $D_c$ are the irreps and the possible charges $c=(q_1,q_2)$ are $(1,1),(1,-1),(-1,1),(-1,-1)$. Since the group is Abelian, tensor products of irreps results in a new irrep that can be obtained by elementwise multiplication in the character table. For example, $D_{({-}1,1)}\otimes D_{(1,{-}1)}=D_{({-}1,{-}1)}$. This product of taking tensor products results in multiplicative operation on charges $c_1\otimes c_2=c_3$ that is often called the fusion rule. Using this fusion rule, we can construct higher order tensors from vectors. Condition (\ref{eqn:SymmetryCondition}) becomes a charge conservation rule\cite{Bauer2011}. In numerics it is possible to identify the nonzero blocks of a given tensor by checking that the block conserves charge. The condition for a nonzero block $B_i$ is

    \begin{align}
      \begin{array}{l}
        \includegraphics{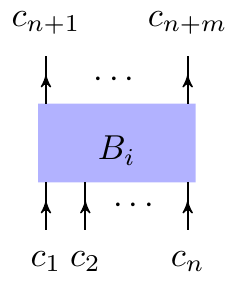}
      \end{array}\hspace{-3mm}\neq0&\iff\prod_{\ell\in\mathrm{In}}c_{\ell}=\prod_{\ell\in\mathrm{Out}}c_{\ell}.\raisetag{1.6\baselineskip}\label{eqn:TensorBlock}
    \end{align}

    Using these block tensors provides us with three main advantages in optimising the network and extracting physical information.

    \subsubsection{Reduction in variational parameters}

    A large number of the blocks are fixed to be zero by the symmetry, so do not have to be stored or manipulated during optimisation. This decrease in variational parameters leads to a decrease in the storage space needed for the network. It also leads to a multiplicative decrease in the computational time ($a\chi^8\to \frac{a}{4}\chi^8$ in the case of $\dd $) allowing either larger $\chi$ to be accessed or a decrease in the overall runtime.
    \subsubsection{Selection of symmetry sector}

    By decomposing the tensors into blocks, any operator applied to the spin chain can be classified according to how it transforms under the symmetry. A operator $O_{q_1,q_2}$ has charge $(q_1, q_2)$ under the $\dd $ symmetry if
    \begin{align}
      S_1OS_1^\dagger=q_1O,\quad S_2OS_2^\dagger=q_2O \,,
    \end{align}
    where $S_1,\,S_2$ are the symmetry generators described in section \ref{sec:CFTs}. Operators carry a unique charge, so operators of the form $O=a_1O_{1,1}+a_2O_{1,-1}$ are forbidden.
    \subsubsection{Nonlocal operators}

    As we know the action of the symmetry on the tensors $\{u,w\}$, we can trivially compute expectation values for a class of highly nonlocal operators, with computational cost which barely exceeds that of local operators\cite{Evenbly2010c}. This is discussed below with respect to scale invariant operators and conformal data.

    \subsection{Optimising a MERA for a spin model}\label{sec:Optimising}

    Given a local, critical spin Hamiltonian $H=\sum_j h_j$, a MERA description of the ground state $\ket{\psi}$ can be generated by optimising the tensors. Here we briefly review the optimisation algorithm, following \inlinecite{Evenbly2011}.

    The MERA for a critical model is built from two sections. First, a number (usually 2-3) of \emph{transitional} layers are used, which are translationally invariant but not scale invariant, with tensors $\{u_\ell,w_\ell\}$. These allow the bond dimension of the remaining (scale invariant) tensors to be chosen independently of the physical dimension of the spins. Generically, the Hamiltonian $H$ will contain RG irrelevant terms, breaking scale invariance. Since these terms become suppressed at larger scales, the transitional layers reduce their effect.

    Above the transitional layers, the MERA is built from a unique pair of tensors $\{u,w\}$, characterising the scale invariant nature of the critical model. We allow the dimension of the upper and lower indices of $u$ to differ; it is not unitary, but Eq.~(\ref{eqn:ucond}) holds. This preserves the essential structure of the MERA, but allows for increased numerical efficiency\cite{Evenbly2010c}. A pair of bond dimensions, $\chi_l$ and $\chi_u$ on the lower and upper indices of $u$ characterise the MERA.

    To initialise the MERA, a pair of bond dimensions $\{\chi_l,\chi_u\}$ is chosen, and random tensors generated. A single transitional layer is initialised to allow $\chi_l$ to differ from the physical dimension $d$. These tensors are then iteratively optimised to minimise the energy $\bra{\psi}H\ket{\psi}=\Tr (H\ketbra{\psi})$ as discussed in \inlinecites{Evenbly2009,Evenbly2011}.

\begin{figure}
      \includegraphics{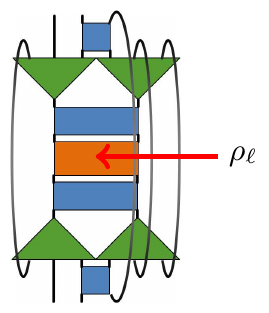}
      \caption{The descending superoperator. This computes the reduced density matrix of the model at level $\ell$ of the RG. Just one of the contractions which must be performed to optimise the MERA tensors. The cost is $\bigO{\chi^8}$, however this can be reduced as discussed in the text.}\label{fig:DescendingSuperoperator}
    \end{figure}

    Optimising the MERA requires many contractions of networks analogous to those shown in Fig.~\ref{fig:DescendingSuperoperator}, where $\rho_\ell$ is the reduced density matrix describing the system on layer $\ell$. This contraction can be computed in time $\bigO{\chi_l^4\chi_u^4}$. In numerical implementation, $\chi_u<\chi_l$ allows for decreased runtime without apparent degradation of the results. By assuming that most of the eigenvalues of the reduced density matrix $\rho$ are small, we can modify the network to improve the scaling as follows\cite{Evenbly2011}. We assume we can choose a projector $P=vv^\dagger$ such that $\rho=P\rho+\epsilon$, where $\epsilon$ is a small error. Then $\bar{\chi}$ is the rank of the projector. In the networks, $\rho$ is replaced with $P\rho$, giving the improved efficiency at the expense of optimising the tensor $v$. Numerical evidence shows that $\bar{\chi}$ may be chosen $\bigO{\chi}$ rather than $\bigO{\chi^2}$, thus providing a scaling of $\bigO{\chi^6\bar{\chi}}$.

    As the optimisation proceeds, the change in the tensors between iterations decreases. Once the network is changing sufficiently slowly, a new transitional layer is added by promoting the lowest level of the scale invariant portion. This process is repeated until it does not produce significant improvement of the ansatz (usually once there are 2-3 transitional layers).

    \subsection{Adiabatic crawling}

    We are simulating families of models with properties that vary continuously along a critical line, and so we can make use of the solution from a converged MERA at one point as a starting point to speed up convergence at a neighbouring point. We call this approach the \emph{adiabatic crawling} technique. First, the MERA is converged from random tensors at a start point somewhere along the line.  The velocity rate of variation of the conformal values (scaling dimensions for relevant operators and central charge)  is defined as $v=C_j-C_{j-1}$, where $C_j$ is the conformal data on iteration $j$. When this falls below some threshold, the point is declared converged, and the algorithm is repeated on the next point, using the previously converged tensors as a start point. If the variation is sufficiently small, the ground state of the new Hamiltonian is very similar to the previous ground state, and few convergence iterations are required. Typically, convergence of the remaining 60-100 points can be completed in a time similar to that required to converge the initial point.

    \subsection{Extracting conformal data from the MERA}

    The properly converged MERA provides a compact approximation to the ground state, and we can extract physical properties of the CFT that describes the spin lattice model at criticality by optimizing a MERA. Here, we describe how to obtain the conformal parameters from the tensors in the MERA, following \inlinecites{Pfeifer2009,Evenbly2010c}.

    One part of the conformal data for a CFT is the spectrum of \emph{scaling dimensions} $\Delta_{\phi}$ associated with a primary field $\phi$. These are the eigenvalues of the rescaling operator in the field theory.  In the MERA, the isometric tensor $w$ performs a rescaling operation on the spin model, so allows us to extract the scaling dimensions\cite{Pfeifer2009}. By finding the operators which are fixed under the one site scaling superoperator
    \begin{align}
      \mathcal{S}(\phi)&=\begin{array}{l}\includegraphics{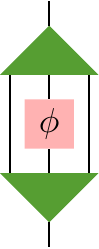}\end{array}=\lambda_\phi\begin{array}{l}\includegraphics{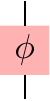}\end{array},
    \end{align}
    we find the scaling operators and their dimensions. The eigenvalues $\lambda_\phi$ of $\mathcal{S}$ are then related to the scaling dimensions via $\Delta_\phi=\log_3(\lambda_\phi)$.

    We can also compute the scaling dimensions for a class of nonlocal scaling operators by making use of the symmetry.  Nonlocal scaling operators take the form
    \begin{align}
      O&=\bigotimes_{j=-\infty}^{N-1}(U_g)_j\otimes o_N,\label{eqn:StringForm}
    \end{align}
    where $U_g$ is a unitary representation of $\mathcal{G}$.
    Since the symmetry operators commute with the MERA tensors, applying one layer to an operator of this form and utilising Eq.~(\ref{eqn:wcond}-\ref{eqn:ucond})
    gives a new operator of the form Eq.~(\ref{eqn:StringForm}).

    Eigenoperators of the \emph{$g$-twisted scaling superoperator}
    \begin{align}
      \mathcal{S}_g(\phi)&=\begin{array}{l}\includegraphics{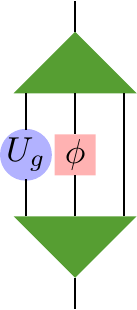}\end{array}=\lambda_\phi\begin{array}{l}\includegraphics{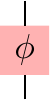}\end{array},\label{eqn:GTwisted}
    \end{align}
    allow extraction of the conformal spectrum of the CFT arising from taking the thermodynamic limit of the spin model with boundary conditions which are twisted by the group element $g$; see~\inlinecite{Evenbly2010c}. Setting $g=e$, $\mathcal{S}$ is recovered, giving the spectrum on periodic boundaries.

    The spectrum of $\mathcal{S}_g$ can be extracted from the converged MERA for negligible additional cost simply by diagonalizing $\mathcal{S}_g$\cite{Pfeifer2009,Evenbly2010c}.

    The central charge of the theory can be extracted from the fixed-point density matrix using the scaling of the entanglement entropy\cite{Pfeifer2009}
    \begin{align}
      S_{\rm critical}&=\frac{c}{3}\log_2(L/a)+k,
    \end{align}
    where $c$ is the central charge of the CFT, $L$ is the block length, $a$ is the lattice scaling (here, $a=1$ by definition) and $k$ is some constant. The one-site reduced density matrix $\rho^{fp}_1$ is obtained by symmetrising over the two ways of tracing out one site of the fixed-point density matrix $\rho^{fp}_2$ (Fig.~\ref{fig:DescendingSuperoperator}). The central charge is then obtained by
    \begin{align}
      c&=3\left(S\left(\rho^{fp}_2\right)-S\left(\rho^{fp}_1\right)\right),
    \end{align}
    where $S$ is the von Neumann entropy of the density matrix.
    
    \section{Numerical results}\label{sec:Results}

    In this section, we present results obtained from a $\dd $ symmetric MERA with $\{\chi_l,\chi_u\}=\{20,12\}$ and $\bar{\chi}=60$ (see Sec.~\ref{sec:Optimising}) applied to the three models described in Sec.~\ref{sec:CFTs}. Three transitional layers are used to ensure scale invariance of the Hamiltonian.  We use the adiabatic crawling technique, starting at $R^2=1.25$ and using a velocity threshold of $v_t=9\times 10^{-5}$.  Both ground state energies and conformal data are presented along the critical lines of the three models given by Eq.~(\ref{eqn:criticalline}). The conformal spectrum is shown to be consistent with the identifications made in Sec.~\ref{sec:CFTs}. We also compare these MERA simulations with the result of exact diagonalisation studies for the models of interest.

    \subsection{Ground state energy}

\begin{figure}[t]
      \begin{tabular}{cc}
        \begin{minipage}{.9\linewidth}
          \includegraphics{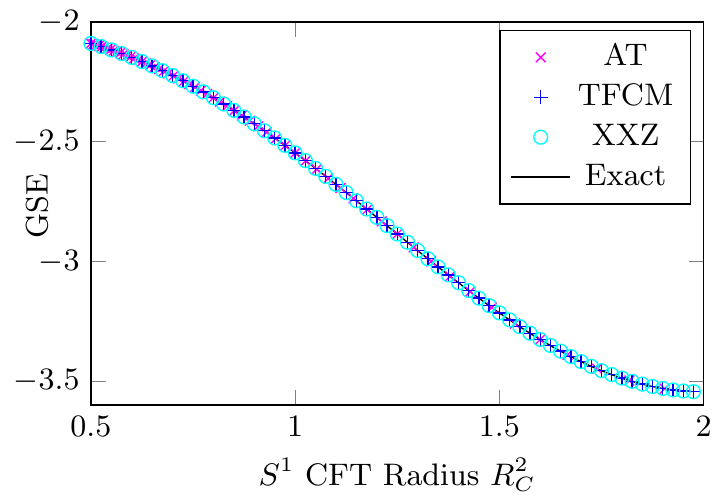}
          \end{minipage}\\
          \begin{minipage}{.9\linewidth}
            \includegraphics{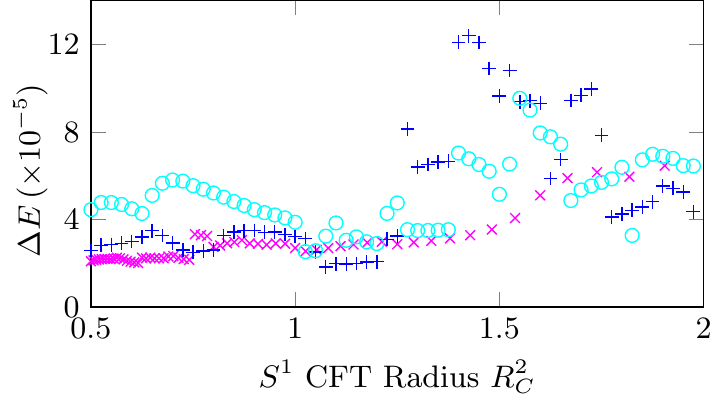}
          \end{minipage}
        \end{tabular}
        \caption{Ground state energies (GSE) extracted from the MERA and their relative errors $\Delta E$. For the bond dimension used here,
        these remain $\bigO{10^{-4}}$ as the Hamiltonian parameter $\lambda$ is varied. The numerical GSE remains an upper bound on the true GSE. Note that the CFT radius $R_C$ is related to the spin model coupling parameter via Eq.~(\ref{eqn:Rc})}\label{fig:GSE}
      \end{figure}

    As a result of the Bethe Ansatz solution for the XXZ model, the ground state energy is known exactly for all three models considered here, and this provides a useful benchmark; see Fig.~\ref{fig:GSE}.
      The relative error in the ground state energy is given by $\Delta E=(E_{\mathrm{exact}}-E_{\mathrm{MERA}})/E_{\mathrm{exact}}$.  The exact solution is obtained by numerically integrating the Bethe Ansatz solution\cite{Alcaraz1988}. The MERA is an explicit wavefunction, and as such, its energy cannot be less than the true ground state energy; thus $\Delta E \geq 0$. The ground state energy obtained using the MERA for all three models is consistent with the exact solution, with relative error positive and of order $10^{-4}$ for the full range of $\lambda$ considered; see Fig.~\ref{fig:GSE}.

      \subsection{Conformal data}

      We now present the conformal data obtained from the MERA for our three models. We show that this data is consistent with the CFTs identified as the thermodynamic description, namely the $\zz $-orbifold boson CFT for the Ashkin-Teller model; and the $S^1$ boson CFT associated with both the XXZ and transverse field cluster models.

      \subsubsection{Central Charge}

\begin{figure}[t]
        \includegraphics{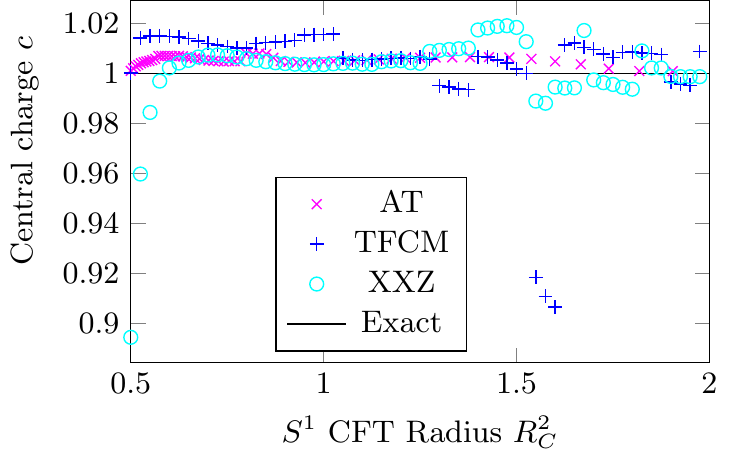}
        \caption{Central charges extracted from MERA simulations. The CFTs associated with these models are expected to be $c=1$ (shown as solid line) for the full range of $R_C^2$ considered here.}\label{fig:c}
      \end{figure}

      One of the pieces of data required to specify a CFT is the central charge. It labels classes of CFT and is identically $c=1$ for all values of $R$ in both the $S^1$ boson and orbifold boson CFTs. The values obtained from the MERA simulations are shown in Fig.~\ref{fig:c}. These are usually within $2\%$ of the expected value $c=1$ for the full range of $R_C^2$.  We note an increased deviation from $c=1$ at the left ($R_C^2 \sim 0.5$) end of the critical line.  Here, fields in the $\dd $ symmetric sector are crossing $\Delta_\phi=2$, the RG relevant/irrelevant threshold. Some of these fields are not symmetric under the $O(2)$ symmetry discussed in section~\ref{sec:CFTs}\cite{Alcaraz1988JPHYS}. Near $R_C^2=1.6$, we also see a large deviation from $c=1$ for the TFCM data. This region occurs close to a crossing of two fields in the $\dd $ symmetric sector of the $S^1$ theory. Again, this crossing does not occur in the $O(2)$ symmetric theory\cite{Alcaraz1988JPHYS}.

      Our simulations approach, but do not include, the end point $\lambda=1,\,(R_C^2=2)$.  At this point, the XXZ model becomes the spin-1/2 Heisenberg model and the line of criticality terminates. As we approach this point, the CFTs have an irrelevant primary operator with a scaling dimension that is decreasing as $R_C$ increases and becomes marginal precisely at $R_C^2=2$. This operator leads to corrections to the scale invariant behaviour that, as has previously been noted\cite{Evenbly2011}, complicate studies of this model using MERA simulations.

      This limiting case could be further investigated in two ways.  One approach would be adding more transitional layers to the MERA, which is expected to move the Hamiltonian towards the RG fixed point.  However, many layers may be required because this term only decays logarithmically. Alternatively, modifying the model by adding finely tuned terms (for example, a next nearest neighbour coupling) could remove the marginally irrelevant operator without changing the continuum limit\cite{Affleck1989,Eggert1996a}.  Both of these approaches are beyond the scope of the current investigation.

      \subsubsection{Scaling dimensions}

\begin{figure}
        \hspace*{-5mm}
        \includegraphics[scale=.9]{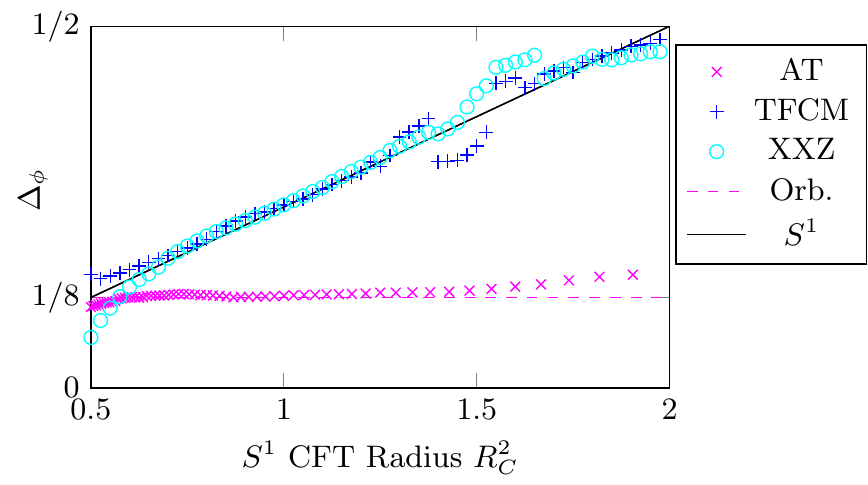}
        \caption{The lowest scaling dimensions ($\Delta_\phi$) of the $(1,-1)$ charge sector of the three models. The XXZ and transverse field cluster models agree well with the $S^1$ boson, whilst the AT model is consistent with the fixed scaling operator from the orbifold theory. Note that the $S^1$ and orbifold CFT radii are related via Eq.~(\ref{eqn:Ro}).}\label{fig:FixedvsMoving}
      \end{figure}

      We now turn to the scaling dimensions $\Delta_\phi$ extracted from the MERA.  These scaling dimensions can be classified according to their $Z_2 \times Z_2$ symmetry sector.
      As the parameters of the models are varied along the lines of criticality, the scaling dimensions of the two CFTs can vary continuously. As noted in section \ref{sec:CFTs}, the orbifold CFT contains a fixed sector of primary fields whose scaling dimensions do not change with $R_O$, whereas for the free boson CFT all scaling dimensions (other than the identity and related operators) will vary.  The behaviour of these scaling dimensions provides a good pointer as to the type of $c=1$ CFT that corresponds to each of the spin models. In Fig.~\ref{fig:FixedvsMoving}, we focus on fields with small scaling dimension in the $(1,-1)$ symmetry sector. Previous evidence\cite{Evenbly2010c,Evenbly2011} indicates that the accuracy of the MERA decreases with increasing scaling dimension, so it is expected that the agreement will be better in this region. The distinction between the CFTs is already evident, and the scaling dimensions obtained from the XXZ and TFCM models are consistent with the $S^1$ theory, and the AT model has the expected fixed scaling dimension present in the orbifold theory.

      \begin{figure*}
        \includegraphics{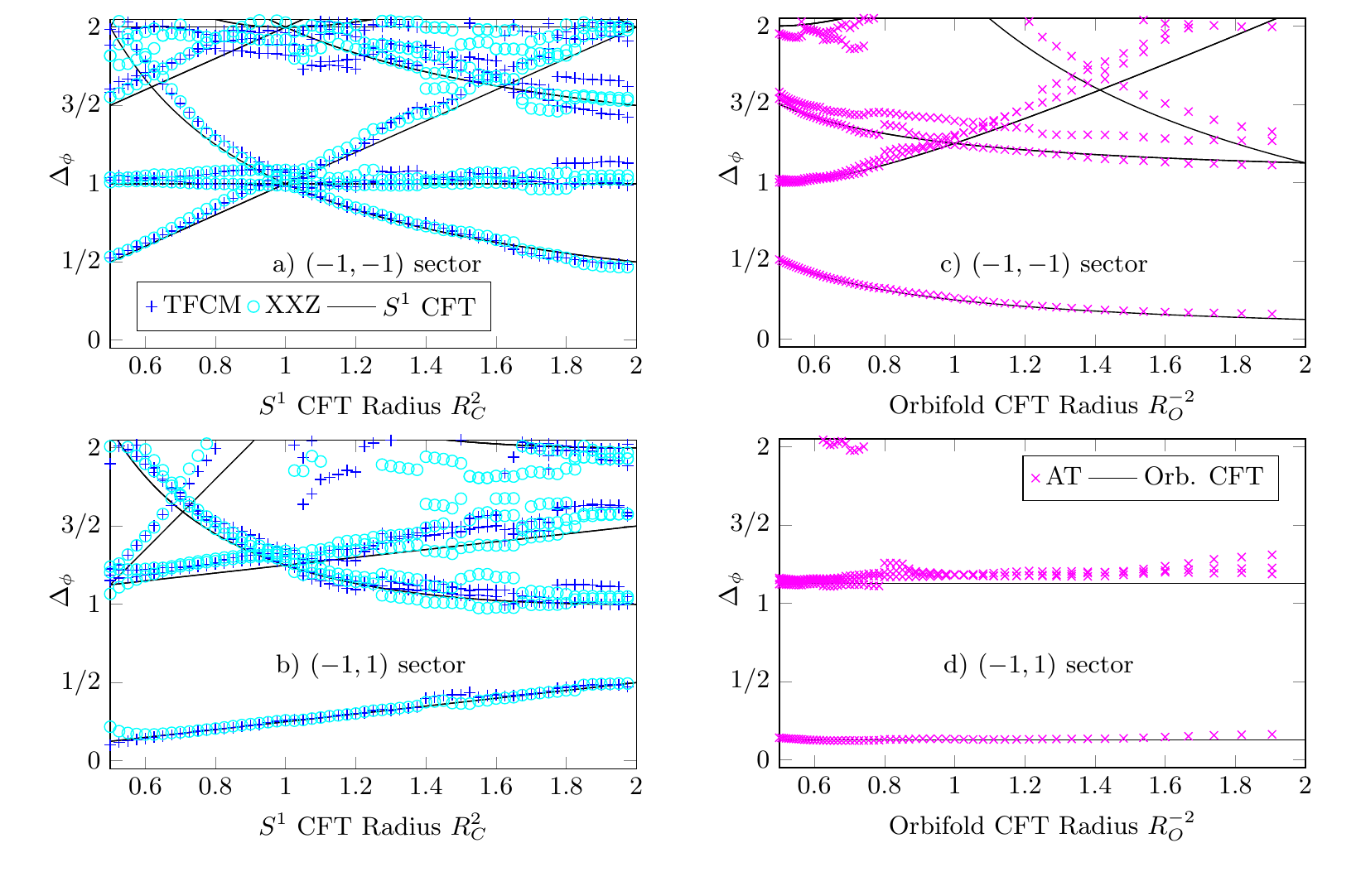}
        \vspace{-8mm}
        \caption{The behaviour expected from the $S^1$ CFT is recovered by the XXZ and TFCM (a,b), whilst AT simulations recover the orbifold behaviour (c,d). The accuracy of the recovered scaling dimensions $\Delta_\phi$ degrades for larger $\Delta_\phi$. Note that the $S^1$ and orbifold CFT radii are related via Eq.~(\ref{eqn:Ro}).}\label{fig:LocalScalingData}
      \end{figure*}

      We next examine the full relevant spectrum of the scaling operators for all three models, up to a maximum value of $\Delta_\phi = 2$.
      All sectors of the XXZ and TFCM models are expected to have the same conformal dimensions; those of the $S^1$ theory. Close agreement is seen in the MERA simulations of the two models (Fig.~\ref{fig:LocalScalingData} a,b), although the accuracy deteriorates for larger scaling dimensions. The conformal field theory has an infinite number of scaling operators, however the one-site scaling superoperator used to obtain $\Delta_\phi$ has a finite number of eigenvalues. This results in only a finite subset of the dimensions being recoverable, with the smallest being most accurately reproduced. In particular, $\Delta_\phi=0$ in the $(1,1)$ sector (Fig.~\ref{fig:FSS}) corresponds to the identity operator. This scaling dimension is recovered perfectly due to the isometric constraint on the tensor $w$.

      The AT model is expected to reproduce the orbifolded theory, with the signature fixed sector being equally split over the $(1,-1)$ and $(-1,1)$ charge sectors.  We see this indeed occurs in Fig.~\ref{fig:LocalScalingData}c, with good agreement for the lower scaling dimensions.

      In the $(1,1)$ sectors (Fig.~\ref{fig:FSS}), we see the deviation increase as we move towards $R_C^2=1/R_O^2=2$. This also occurs in the XXZ and TFCM models. Recall that, at this endpoint, all three models are unitarily equivalent to the Heisenberg model.
      \subsection{Exact diagonalisation of the quantum Ashkin-Teller model}

      \begin{figure*}
        \includegraphics{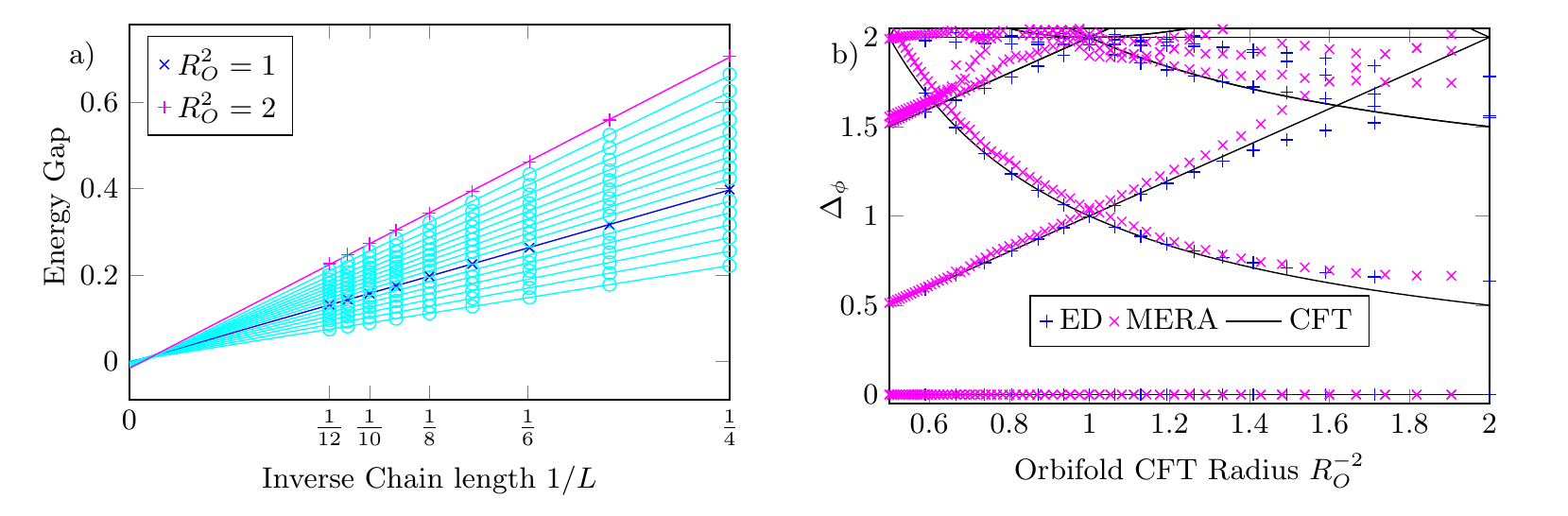}
        \caption{Results of an exact diagonalisation study of the Ashkin-Teller model in the symmetric $(1,1)$ sector. a) The energy gap for a range of values of $R_O$ and system sizes $L$. The gap closes as we approach the thermodynamic limit. The gap is plotted for various $1/R_O^2$  values; in blue (\textcolor{blue}{$\times$}) is the point $1/R_O^2=1$ and in magenta (\textcolor{magenta}{$+$}) the point $1/R_O^2=2$. Lines represent a linear fit to the ED data. b) A comparison of the finite-size scaling results for $L=12$ with the MERA and the CFT, for the $\dd $ symmetric sector.}\label{fig:FSS}
      \end{figure*}

      It is well known that the low energy spectrum of finite size spin chains should correspond to the spectrum of primary and descendant fields of the associated CFT~\cite{DiFrancesco1997}. As such the results of exact diagonalisation (ED) with periodic boundary conditions can be directly compared with those of both CFT and the MERA simulations. For a review of the ED technique, see for example \inlinecite{EDBook}.

      As a further test of the MERA simulations, we performed ED calculations for the Ashkin-Teller model with periodic boundary conditions for chains up to $L=12$ in length.  At this length, finite-size effects are still present but we are nevertheless close to full convergence. As can be seen in Fig.~\ref{fig:FSS}a, the gap is closing like $1/L$, as is expected of a critical model~\cite{DiFrancesco1997}.

      The spectrum of the quantum spin chain can be related to that of the CFT once the  ground state energy ($E_0$) and the overall normalisation of the Hamiltonian are chosen appropriately\cite{Reugg1985}. In our problem we need to choose a normalisation of H for each λ and we achieve this simply by rescaling the gap ($E_1-E_0$) of the spin model. Then the rest of the conformal spectrum can be inferred from the spectrum of the spin chain as follows
      \begin{align}
        \Delta_k&=L\times\alpha(\lambda)(E_k-E_0),
      \end{align}
      where, anticipating the orbifold CFT, the $\alpha(\lambda)$ is chosen to fix the gap $\Delta_1=1/8$. This value corresponds to the scaling dimension of the operators with fixed scaling dimension in the $\left(\pm,\mp\right)$ sectors (Fig.~\ref{fig:LocalScalingData}d). Thus the first two scaling dimensions of the CFT are correct as a result of our choice of normalisation and the zero of energy. Obtaining the remaining levels represents a confirmation of the CFT identification. The same approach was taken in previous studies of the quantum Ashkin-Teller model\cite{Baake1987,Baake1987a}, although there an analytic form for $\alpha$ was proposed.

      We see good qualitative agreement between the two methods. A selected $\dd $ symmetry sector is presented in Fig.~\ref{fig:FSS} b; we see comparable agreement for all other sectors. The agreement between the two methods is extremely good for lower scaling dimensions but becomes less so for higher ones above $\Delta_\phi \approx 1.5$. The ED and CFT data match very precisely at the point $1/R_O^2=1$ for all scaling dimensions up to $\Delta_\phi \simeq 2$, where the model is equivalent to two uncoupled critical Ising models. Hence, away from the decoupling point we interpret the decrease in precision with increasing scaling dimension as a finite-size effect rather than a result of the Lanczos-based ED algorithm.

      It is interesting to note that close to the point $1/R_O^2=2$ the MERA and the ED results agree with each other rather better than either one agrees with the CFT values. This is the limit in which the finite size corrections to the eigenvalues for the spin chain, which are well understood for ED\cite{Christe}, increase in size due to an irrelevant primary field whose scaling dimension is approaching 2. It would be an interesting avenue for further study to understand whether the errors in scaling dimensions obtained from MERA simulations can be understood in a similar way to those of ED.

      We note that finite-size scaling has previously been performed for the Ashkin-Teller chain, in order to identify the primary fields for periodic and anti-periodic boundary conditions \cite{Gehlen1987,Baake1987,Baake1987a}. To the best of our knowledge, our ED calculations use a system size equal to the largest used in all previous studies\cite{Yamanaka1995}. Here we have extended those calculations over a larger range of $1/R_O^2$ as well as extracted scaling dimensions for all RG relevant fields.

      \subsection{Nonlocal operators and twisted boundaries}

      Another class of scaling operators present in the symmetric MERA are the nonlocal operators involving a half infinite string of symmetry operators terminating in a local operator. The scaling dimensions of these operators in the CFT arise in ED studies of the corresponding spin model with boundary conditions twisted by a group element. Recall that the scaling dimension of these nonlocal operators can be obtained from MERA simulations at negligible additional computational cost.

\begin{figure}
        \includegraphics{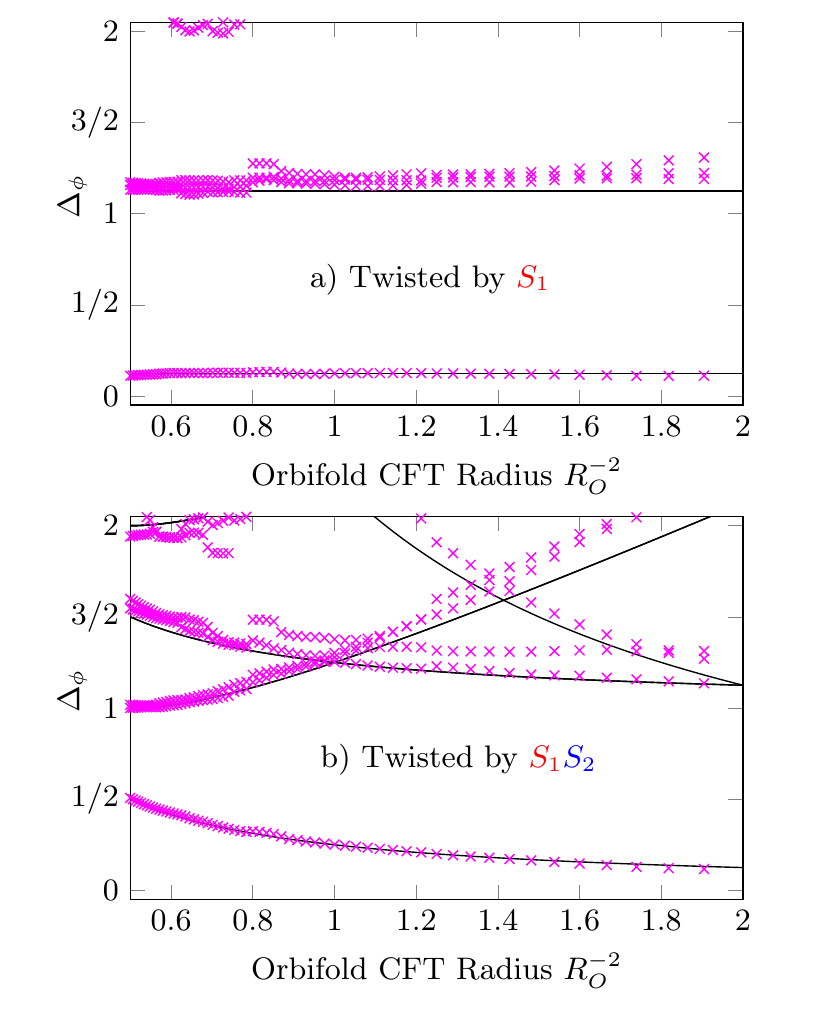}
        \caption{Scaling dimensions of nonlocal operators extracted from MERA simulations. These correspond to the spectrum of spin chains with boundary conditions twisted by the group elements defined in Eq.~(\ref{eqn:ATSymmetryOperators}).} \label{fig:ATboundaries}
      \end{figure}

      In Fig.~\ref{fig:ATboundaries}, we show these nonlocal scaling operators in the $(1,1)$ symmetry sector for two kind of twist (corresponding to the two of the group elements of $\dd $) obtained from the MERA simulation of the $\dd $ symmetric AT model. Note that Fig.~\ref{fig:ATboundaries} and Fig.~\ref{fig:LocalScalingData} show different data. The orbifold CFT is such that the local (untwisted) spectrum in the four symmetry sectors is equivalent to the twisted spectrum in the $(1,1)$ sector.  This equivalence is apparent by comparing Fig.~\ref{fig:LocalScalingData}c (d) with Fig.~\ref{fig:ATboundaries}b (a).
      The deviation of the scaling dimensions of these nonlocal fields from the CFT expectation appears to be comparable to that of the local fields.

      \subsection{Observations and Comments}
      \begin{figure}
        \includegraphics{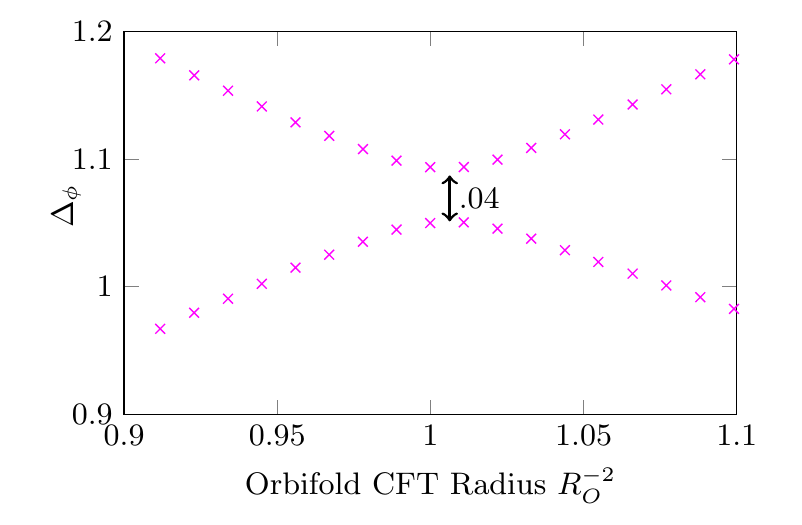}
        \caption{Only the $\dd $ subgroup of the $D_4$ symmetry of the Ashkin-Teller model is enforced in the MERA. This can lead to coupling between scaling operators which one would expect to be forbidden, in turn leading to avoided crossings in the conformal dimensions. This data has $\chi_l=20=\bar{\chi}/4\,,\chi_u=16$ and 4 transitional layers. }\label{fig:AvoidedCrossing}
      \end{figure}

      All the models considered here have a larger on-site symmetry than the enforced $\dd $. In particular, the AT model is invariant under swapping $\rs{}{}\leftrightarrow\bt{}{}$, leading to a nonabelian $D_4$ invariance.
      The symmetry group of the staggered XXZ model contains an $SO(2)$ symmetry, for $SO(2)$ rotations of all spins about the $z$-axis, and a $\zz$ symmetry corresponding to $\pi$-flips around the $x$-axis, just as the free boson CFT as described in Sec.~\ref{sec:CFTs}. This maps to an on-site symmetry in TFCM, but becomes a nonlocal symmetry in the Ashkin-Teller model unless the $SO(2)$ angle is $\pi$ (in which case this symmetry reduces to the $\dd $ symmetry we have enforced).

      Enforcing only a proper subgroup of the full symmetry means that the results of the MERA simulations do not respect the full symmetry of the model. We note that this leads to avoided crossings in the conformal dimensions extracted from the simulations, as seen in Fig.~\ref{fig:AvoidedCrossing}. The two operators in this plot transform like different representations of $D_4$, and so are forbidden to mix. Under the $\dd $ subgroup used in our simulations, such a mixing is allowed, and is indeed observed. We expect this gap to close as the approximation is improved, leading to an approximately enforced $D_4$.

      We also note that under the full conformal symmetry, all primary fields transform like different irreps, meaning they are all uncoupled.  As the MERA does not incorporate the full (continuous) conformal symmetry in its structure, any MERA-based numerical method cannot keep all such fields from mixing, and as such we expect avoided crossings to be observed even if the full on-site group is enforced.
      
We also note that the choice of starting point for our adiabatic crawling method appears to have an effect on the accuracy of the resulting converged MERA. Specifically, at values of $R_C$ or $R_O$ where previously irrelevant operators cross over to become relevant, increased errors arise that slowly die away as the simulation proceeds. A similar situation occurs when the scaling dimensions of two relevant operators cross as $R_C$ and $R_O$ are varied. Crawling from multiple points, and stopping at crossings and when new operators become relevant may reduce this behaviour.
      
      \subsubsection{Choice of $\chi$}
      In this work, we have made a choice of the bond dimensions $\{\chi_l,\chi_u\}=\{20,12\}$ and $\bar{\chi}=60$ but we have investigated a range of $\chi$ values both higher and lower than this. These values were set by our available computational resources but we do not expect the essential conclusions to be altered by larger values of $\chi$. The scaling of the error in the MERA energy was investigated in \inlinecite{Evenbly2011} and elsewhere and is one of the few quantitative methods of studying convergence of MERA simulations reported in the literature. As discussed above, we can compare the energy of our MERA simulation with the exact result available from the Bethe ansatz solution and we also observe that for increased  $\chi$ the error in the energy is reduced, as one would expect. Indeed the error in Fig.~\ref{fig:GSE} is essentially the same size as reported in \inlinecite{Evenbly2011} for the staggered Heisenberg and XX models and the same value of $\chi$  so we believe that our simulations have comparable accuracy to other implementations of MERA with comparable values of $\chi$.
      
      We have also qualitatively studied the convergence of the smaller scaling dimensions and central charge to the values predicted by CFT. The error in both decreases slowly with increasing $\chi$, with the smallest scaling dimensions being recovered more accurately than those higher.
      
The qualitative features we see in the spectra of scaling dimensions are also robust and observed over a range of choices of $\chi$. For example the avoided crossings remain present even when increasing to $\{\chi_l,\chi_u,\bar{\chi}\}=\{20,20,80\}$. Finally, we note that the deviation between the results obtained from the MERA and the CFT result as we approach the endpoint of the critical line corresponding to $R_C^2=1/R_O^2=2$ continues to persist even as $\chi$ is increased.
      
      \section{Summary and discussion}\label{sec:Conclusions}

      In this work, we have used independently-developed scale invariant $\dd $ symmetric MERA code to investigate critical quantum spin lattice models with this symmetry.  Specifically, we have used this to simulate the staggered XXZ and transverse field cluster models, extracting conformal data consistent with the $S^1$ boson conformal field theory being the thermodynamic description of both spin chains. This $c=1$ theory has a parameter $R_C$ which can be varied, leading to continuously varying critical exponents, a behaviour which has been replicated in the MERA.
      In addition, we have extracted conformal data for the Ashkin-Teller spin chain, identifying the $\zz $-orbifold boson as the appropriate CFT. Once again, the behaviour of the scaling fields with the parameter $R_O$ has been recovered, including the fixed sectors which are signatures of this theory.

      We have introduced a crawling method that allows for efficient optimisation of MERA along the critical lines.  By examining the symmetries of the three models, we have identified some of the limitations of this technique, and how enforcing only a subgroup of the full symmetry is revealed in the conformal dimensions. We have also identified similarities between the numerical results of MERA and exact diagonalisation, in the behaviour of errors at different regions along the critical line.

      We note that the staggered XXZ and transverse field cluster models possess phases with nontrivial $\dd $ symmetry protected topological order. Such phases support gapless edge modes that are protected by the symmetry, a property not shared by SPT trivial phases\cite{ChenGuLiuWen}. Understanding the critical theory occurring at the transition between the trivial and SPTO phases may provide insight into the fate of these edge states at the phase transition and their properties within the phase\cite{Chen2013248}. We have seen how tuning parameters in the spin models leads to continuous variation in the critical theory, despite the fact that we are investigating the transition from a single SPT phase to a single trivial phase.

      Recent developments of MERA allow for the incorporation of conformal defects, including interfaces and boundaries\cite{Evenbly2010d,Evenbly2014}. These numerical tools offer the possibility to study the gapless edge modes via the `domain wall boundary conditions' described in \inlinecite{Chen2013248} as well as interfaces between different SPT phases\cite{Lu2014GSE}.

      \begin{acknowledgements}
        We thank Guifre Vidal and Sukhwinder Singh for useful discussions.  We acknowledge support from the ARC via project number DP130103715 and the Centre of Excellence in Engineered Quantum Systems (EQuS), project number CE110001013.
      \end{acknowledgements}

    \end{document}